\documentclass[a4paper]{jpconf}
\usepackage{graphicx}

\begin{document}

\title{Rotating NSs/QSs and recent astrophysical observations}
\author{Ang Li}
\address{Department of Astronomy, Xiamen University, Xiamen, Fujian 361005, China}
\ead{liang@xmu.edu.cn}

\author{Jianmin Dong}
\address{Institute of Modern Physics, Chinese Academy of Sciences, Lanzhou 730000, China}
\ead{dongjm07@impcas.ac.cn}

\begin{abstract}
Both fast and slow configurations of rotating neutron stars (NSs) are studied with the recently-constructed unified NS EoSs. The calculations for pure quark stars (QSs) and hybrid stars (HSs) are also done, using several updated quark matter EoSs and Gibbs construction for obtaining hadron-quark mixed phase. All three types of EoSs fulfill the recent 2-solar-mass constrain. By confronting the glitch observations with the theoretical calculations for the crustal moment of inertia (MoI), we find that the glitch crisis is still present in Vela-like pulsars. An upcoming accurate MoI measurement (eg., a possible 10\% accuracy for pulsar PSR J0737-3039A) could distinguish QSs from NSs, since the MoIs of QSs are generally $> \sim 1.5 $ times larger than NSs and HSs, no matter the compactness and the mass of the stars. Using tabulated EoSs, we compute stationary and equilibrium sequences of rapidly rotating, relativistic stars in general relativity from the well-tested $rns$ code, assuming the matter comprising the star to be a perfect fluid. All three observed properties of the short gamma-ray bursts (SGRBs) internal plateaus sample are simulated using the rotating configurations of NSs/QSs as basic input. We finally argue that for some characteristic SGRBs, the post-merger products of NS-NS mergers are probably supramassive QSs rather than NSs, and NS-NS mergers are a plausible location for quark deconfinement and the formation of QSs.
\end{abstract}
\section{Introduction}

Neutron stars (NSs) are by far one of the most interesting observational objects, since many mysteries remain on them due to their complexity. Future observations with advanced telescopes such as SKA, Advanced LIGO/VIRGO, AXTAR, FAST, NICER, LOFT, and eXTP, will certainly improve our current knowledge of such stellar objects, by providing precise measurements of their masses, radii, spin frequencies, glitches, surface temperatures, luminosities, magnetic strengthes, etc. Theoretically, since no first-principle theory can be used due to the non-perturbative nature of quantum chromodynamics at finite chemical density, their inner structures are still unclear. There is also the hypothesis of pure quark stars (QSs) and their transitions to NSs are both intriguing and important open problems. Therefore, for the study of NSs/QSs, particular attention should be paid on combining realistic enough theoretical models with accumulated observations. The present contribution is along this line.

For this purpose, we select unified NS EoSs that satisfy up-to-date experimental constraints from both nuclear physics~\cite{flow} and astrophysics~\cite{2mass10,2mass13,2mass16}, based on modern nuclear many-body theories, including microscopic methods (e.g.~BCPM~\cite{bcpm}) starting from a realistic nucleon-nucleon two-body force (usually accompanied with a nucleonic three body force) and phenomenological ones (e.g.~BSk20, BSk21~\cite{bsk}, Shen~\cite{shen}), starting from a nuclear effective force with parameters fitted to finite nuclei experiments and/or nuclear saturation properties; we also find typical parameter sets for QS EoSs in developed confinement density-dependent mass (CDDM) model, labelled as CIDDM, CDDM1, CDDM2~(see \cite{ang16prd} and references therein). Hadron-quark mixed phase is obtained by Gibbs construction~\cite{anghybrid2,anghybrid1} and the calculations of hybrid stars (HSs) are done as well. All three types of EoSs fulfill the recent 2-solar-mass constrain~\cite{2mass10,2mass13,2mass16}. For simplicity, several exotic phases possible in NSs' cores are not included here, for example, hyperons~(see, e.g.,~\cite{angy}), kaon meson condensation~(see, e.g.,~\cite{angk}), Delta excitation~(see, e.g.,~\cite{Li16d}). In the following, we introduce briefly the theoretical framework and discussions of our results in Sects. $2 - 4$, before providing the summary and future perspectives in Sect.~5. More details can be found in Refs.~\cite{ang16prd,ang16apjs,ang17} and references therein.

\section{Structures of the Vela pulsar and the glitch crisis}

The Vela pulsar (PSR B0833-45) is among the most studied pulsars, which are usually regarded as highly-magnetized, rapidly rotating NSs. By far there is an accurate determination of several pulsar properties, such as the distance, the spin period ($89.33$ milliseconds), and 17 quasi-periodically produced glitches (sudden increases in spin frequency). Those observations will be used here for a detailed study of its global properties and inner structures. In particular, the ``glitch crisis'' problem (see eg:~\cite{crisis12,crisis13}), relating to the moment of inertia (MoI) of the stellar crust(see eg:~\cite{Link99,li15}), has been a great debate recently and can be discussed as well.

Assuming the pulsar mass ranging from $1.0M_{\odot}$ to $2.0M_{\odot}$, we calculate the detailed structures of this pulsar, namely the central density, the core/crust radii, the core/crustal mass, the core/crustal thickness, the total/crustal MoI. They are collected in Tables $2 - 4$ of Ref.~\cite{ang16apjs}. We show here the results from BSk20 EoS as an example.

Furthermore, we analyse the ``glitch crisis'' problem based on the calculated MoI results. In Fig.~1, the fractional MoIs $I_{\rm c}/I$ are plotted as a function of the stellar mass for three unified NS EoSs (BCPM, BSk20, BSk21), to be compared with one matched EoS ``BHF + NV + BPS''. We first notice that for the previous glitch constraint $I_{\rm c}/I \geq 1.6\%$, the mass of the Vela pulsar should only be smaller than 1.8M$_{\odot}$ based on selected NS EoS models here. This could be easily fulfilled for an isolated pulsar. However, the discovered crustal entrainment shifts the constrain to around $I_{\rm c}/I \geq 7\%$, which could hardly be fulfilled by a pulsar heavier than $1.0 M_{\odot}$, except a narrow range of $1.0 \sim 1.06 M_{\odot}$ in the matched ``BHF + NV + BPS'' EoS case. Actually the predicted fractional moments of inertia based on three unified NS EoSs (BCPM, BSk21, BSk20) are no more than $6.48 \%$. Therefore, we argue that the crisis could be still present in the standard two-component model.

\begin{figure}[h]
\includegraphics[width=20pc]{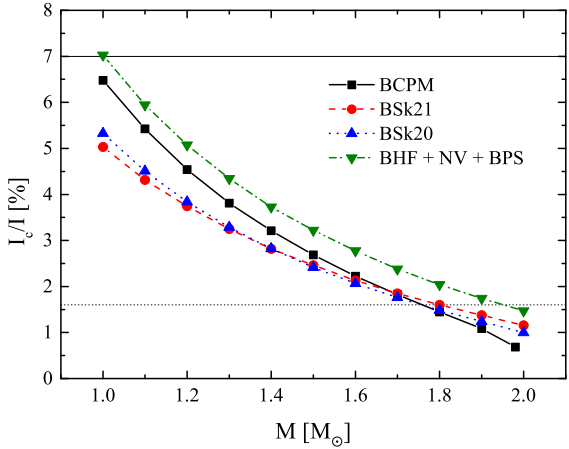}\hspace{2pc}%
\begin{minipage}[b]{14pc}\caption{Fractional moments of inertia $I_{\rm c}/I$ as a function of the stellar mass, with three unified NS EoSs (BCPM, BSk21, BSk20) and one of the matched EoSs (BHF + NV +BPS). Taken from Ref.~\cite{ang16apjs} and see there for details.}
\end{minipage}
\end{figure}

\begin{table}[!htb]
\begin{center}
\caption{Predictions for the properties of the Vela pulsar (PSR B0833-45) with a spin period of 89.33 milliseconds, taking the pulsar mass ranging from $1.0M_{\odot}$ to $2.0M_{\odot}$. The cental densities are in unites of fm$^{-3}$, the masses in solar masses $M_{\odot}$ (except the mass of the outer crust $M_{\rm ocrust}$ in $10^{-5}M_{\odot}$), the radii in kilometers, the total moments of inertia in $10^{45}$g cm$^2$, and the fractional moments of inertia in percentage. Calculations are done with the BSk20 NS EoS, which is based on the widely-used Skyrme nuclear energy density functional, and the parameters were fitted to reproduce with high accuracy almost all known nuclear masses, and to various physical conditions including the nuclear matter EoS from microscopic calculations. Taken from Ref.~\cite{ang16apjs}  and see there for details and more results on other unified EoSs (eg: BCPM, BSk21).}
\begin{tabular}{cc ccc cccc cc}
\hline\hline
            {Mass~}   {Cent.}
            &\multicolumn{3}{c}{Mass }
            &\multicolumn{4}{c}{Radius}
            &\multicolumn{2}{c}{Moment of inertia} \\
            &{core}&{icrust} &{ocrust}
            &{total}&{core} &{icrust} &{ocrust}
            &{total}&{fraction} \\
\hline
             {1.0~~} {0.328}
            &{0.97}&{0.025} &{5.57}
            &{12.50}&{10.93} &{0.84} &{0.73}
            &{1.001} &{5.03}   \\
             {1.1~~} {0.346}
            &{1.08}&{0.023} &{5.08}
            &{12.54}&{11.13} &{0.76} &{0.65}
            &{1.154} &{4.32}   \\
             {1.2~~} {0.365}
            &{1.18}&{0.022} &{4.68}
            &{12.58}&{11.30} &{0.69} &{0.58}
            &{1.315} &{3.75}   \\
             {1.3~~} {0.384}
            &{1.28}&{0.021} &{4.25}
            &{12.60}&{11.44} &{0.63} &{0.53}
            &{1.482} &{3.25}   \\
             {1.4~~} {0.405}
            &{1.38}&{0.019} &{3.93}
            &{12.62}&{11.57} &{0.57} &{0.48}
            &{1.657} &{2.82}   \\
             {1.5~~} {0.428}
            &{1.48}&{0.018} &{3.57}
            &{12.62}&{11.67} &{0.52} &{0.43}
            &{1.839} &{2.46}   \\
             {1.6~~} {0.453}
            &{1.58}&{0.016} &{3.27}
            &{12.61}&{11.74} &{0.48} &{0.39}
            &{2.026} &{2.13}   \\
             {1.7~~} {0.480}
            &{1.68}&{0.015} &{2.93}
            &{12.58}&{11.79} &{0.43} &{0.35}
            &{2.220} &{1.85}   \\                                            {1.8~~} {0.511}
            &{1.79}&{0.014} &{2.63}
            &{12.52}&{11.82} &{0.39} &{0.31}
            &{2.420} &{1.60}   \\
             {1.9~~} {0.548}
            &{1.89}&{0.012} &{2.34}
            &{12.44}&{11.81} &{0.35} &{0.28}
            &{2.624} &{1.38}   \\
             {2.0~~} {0.593}
            &{1.99}&{0.011} &{2.05}
            &{12.32}&{11.76} &{0.31} &{0.25}
            &{2.833} &{1.16}   \\
\hline\hline

\end{tabular}
\end{center}
\end{table}

\section{NSs or QSs? MoI as a probe}

The coexistence of NSs and QSs has been a great puzzle. Recently, MoI has be suggested to be measured rather accurately for pulsar PSR J0737-3039A, a 10\% accuracy (or a constraint on $R\sim5\%$), using the double pulsar system within the next 20 years. This would significantly constrain the EoS for dense stellar matter and may serve as a tool to justify/discriminate NSs/QSs, see references~\cite{MoI04apj,MoI05mn,MoI05apj,MoI13mn,MoI07npa,angq} and references therein for more details.

In Ref.~\cite{anghybrid2}, we developed our previous study~\cite{anghybrid1} of the transition to deconfined quark phase in NSs, by including the
interaction in the quark EoS to the leading order in the perturbative expansion within the CDDM model. Using the Gibbs conditions the hadron-quark mixed phase is constructed matching
the latter with the hadron EoS derived from the microscopic Brueckner theory. High-mass HSs consistent with 2-solar-mass constrain are possible with mixed-phase core with typical parameter sets. Together with EoS models based on the same theories for pure NSs and QSs, respectively, we are ready to evaluate the MoIs of three types of compact stars, and study whether or not the upcoming MoI constrain might serve as probe for high-density phase of stellar matter. In addition, previous studies (eg:~\cite{MoI04apj,MoI05mn}) found large spread in the results for self-bound QSs. The reason might be relatively free parameters in QS models which determine the surface density of strange quark matter at zero pressure. This situation could be largely improved by the new 2-solar-mass constrain used in the present study.

Fig.~2 present the stars' gravitational masses as functions of central density for these EoSs models (NSs in solid black line, SSs in solid red and blue lines, and HSs in dash-dotted blue line), together with two recent massive stars (PSR J1614-2230~\cite{2mass10,2mass13} and PSR J0348+0432~\cite{2mass16}) whose masses are accurately measured. The horizontal line represents the pulsar mass $1.337M_{\odot}$ of PSR J0737-3039A. The calculated MoI of QSs are generally $> \sim 1.5 $ times larger than those of NSs and HSs, no matter the compactness and the mass of the stars. This is consistent with the calculations in Ref.~\cite{MoI13mn}. Fig.~3 shows the scaled MoIs as a function of stellar mass, in comparison with an upcoming accurate MoI measurement (indicated by a shadowed area like Ref.~\cite{MoI05apj}). One see that such measurement could have the ability to distinguish QSs from NSs/HSs. More model calculations should be done to justify this argument as a model-independent one.

\begin{figure}[h]
\begin{minipage}{16pc}
\includegraphics[width=16pc]{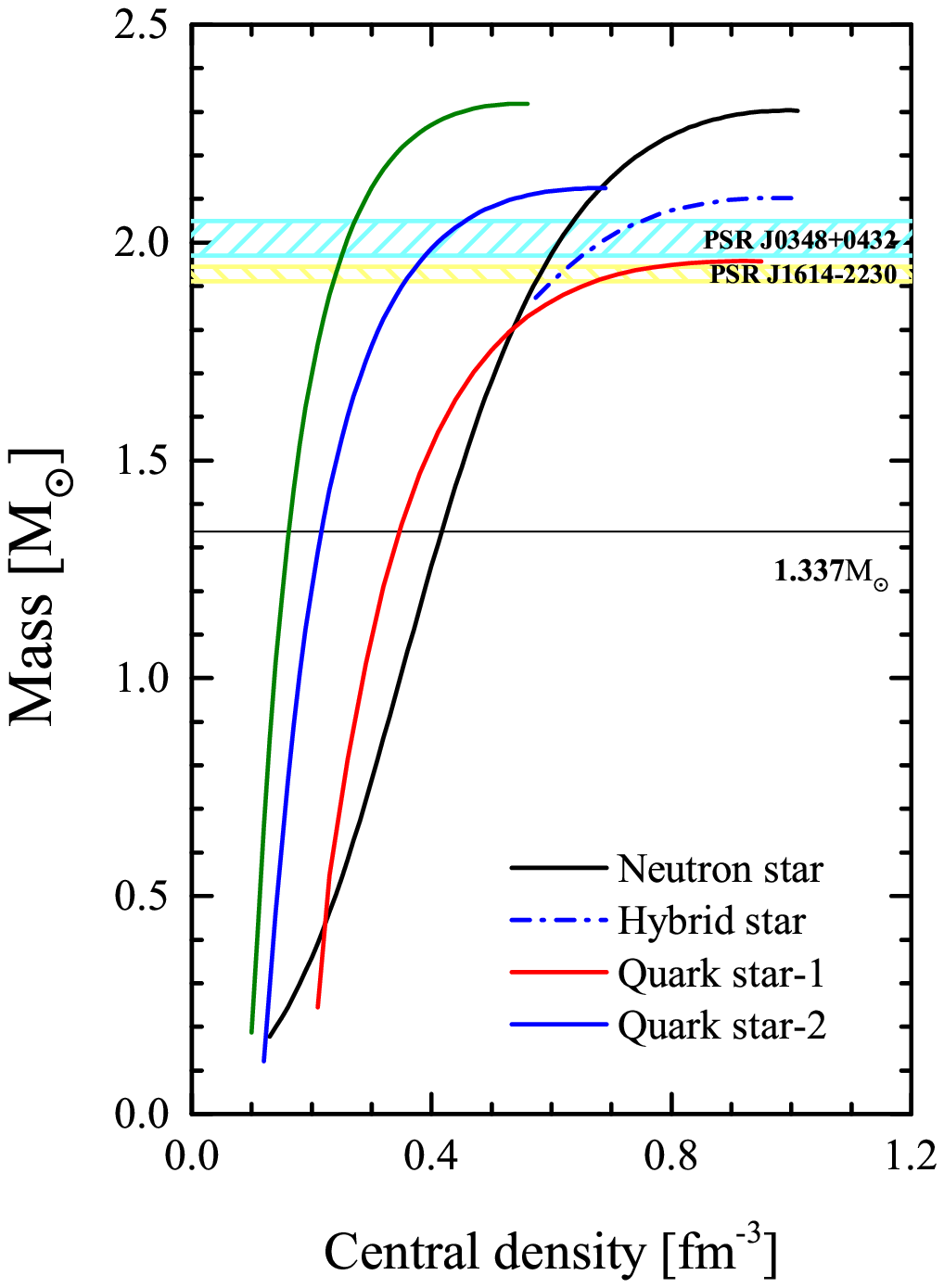}
\caption{Gravitational masses as functions of the stars' central densities for NSs, HSs and QSs, together with two recently-measured massive stars. The horizontal line represents the pulsar mass $1.337M_{\odot}$ of PSR J0737-3039A. Taken from Ref.~\cite{ang17} and see there for details.}
\end{minipage}\hspace{2pc}%
\begin{minipage}{16pc}
\includegraphics[width=16pc]{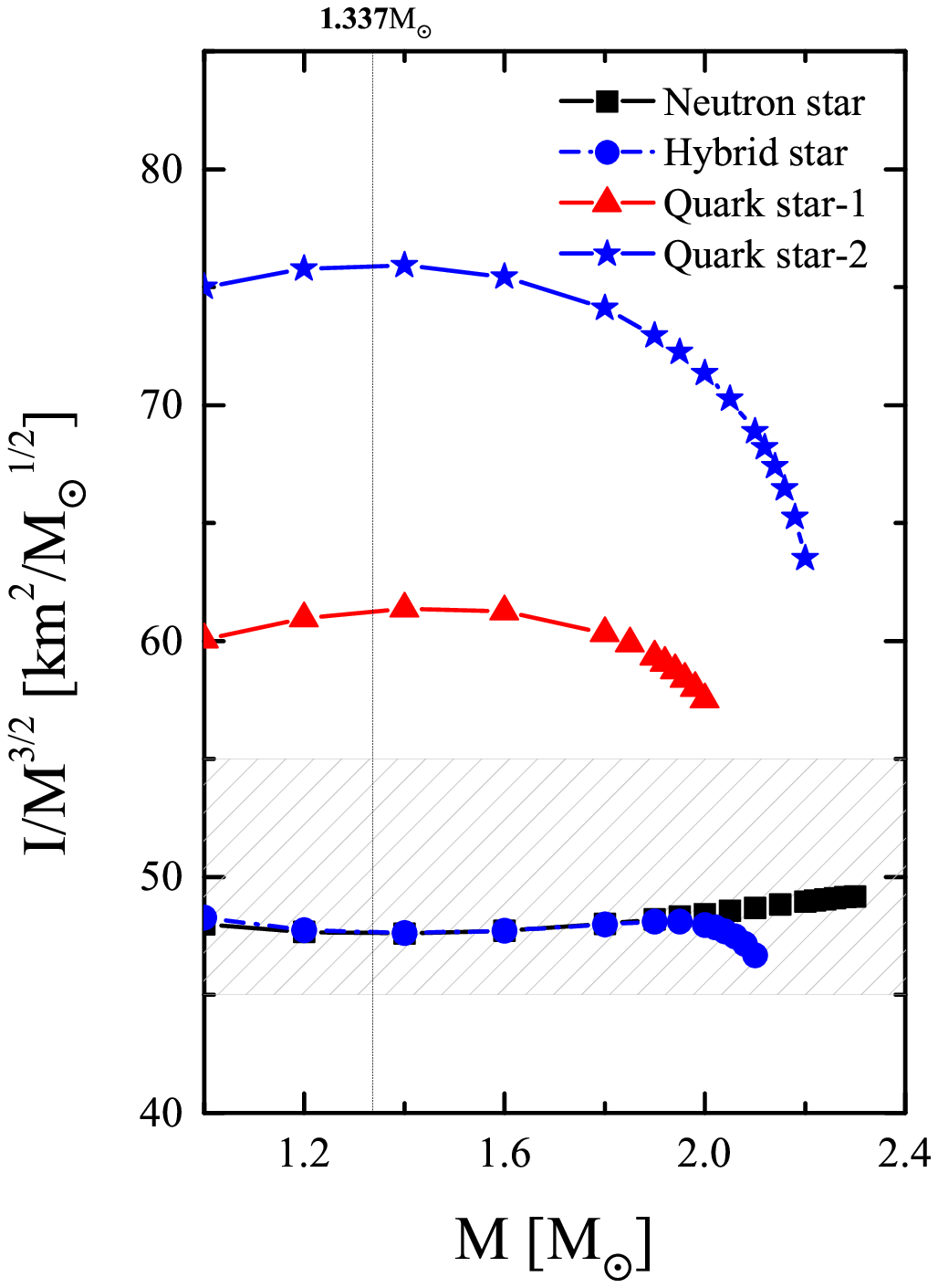}
\caption{Scaled MoI as a function of stellar mass, based on EoS models in Fig.~2. Shadowed area indicates an upcoming accurate MoI measurement~\cite{MoI05apj}. The vertical line represents the pulsar mass $1.337M_{\odot}$ of PSR J0737-3039A. Taken from Ref.~\cite{ang17} and see there for details.}
\end{minipage}
\end{figure}

\section{Internal X-ray plateau in short GRBs: Signature of supramassive fast-rotating QSs?}

\begin{figure}[h]
\includegraphics[width=20pc]{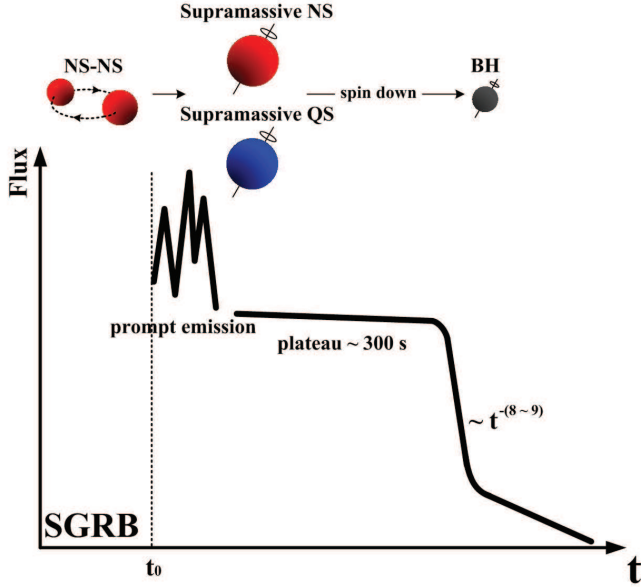}\hspace{2pc}%
\begin{minipage}[b]{14pc}
\caption{Illustration of the ``internal plateau'' observation and the NS/QS central engine model in SGRBs.}\end{minipage}
\end{figure}

Short gamma-ray bursts (SGRBs) are generally believed to originate from the mergers of two NSs or one NS and one black hole (NS-BH). The traditional view is that NS-NS mergers produce a BH
promptly or less than 1 second after the merger, and accretion
of remaining debris into the BH launches a relativistic jet that
powers the SGRB. Such an engine is also naturally expected
within NS-BH merger systems. $Swift$ observations of
SGRBs, on the other hand, showed extended central engine
activities in the early X-ray afterglow phase, in particular the so-called ¡°internal plateau¡±,  characterized by a nearly flat light curve plateau extending to $\sim$ 300 seconds followed by a rapid $t^{-(8 \sim 9)}$ decay, as shown in Fig.~4.

Since it is very difficult for a BH engine to power such a plateau, one attractive interpretation is that NS-NS mergers produce a rapidly-spinning, supramassive NS rather than a BH, with the rapid decay phase signify the epoch when the supramassive NS collapses to a BH after the NS spins down due to dipole radiation or gravitational wave radiation. Whether the current modelling of a NS could reproduce reasonably all three observed quantities (the break time $t_{\rm b}$, the break time luminosity $L_{\rm b}$ and the total energy in the electromagnetic channel $E_{\rm total}$) is crucially related to the underlying EoS of dense nuclear matter and the calculated rotating configurations.

\begin{table*}
\tabcolsep 1pt
\caption{NS/QS EoSs investigated in this study. Here $P_{\rm K}$, $I_{\rm K, max}$ are the Keplerian spin limit and the corresponding maximum moment of inertia, respectively; $M_{\rm TOV}$, $R_{\rm eq}$ are the static gravitational maximum mass and the corresponding equatorial radius, respectively; $\alpha, \beta$ are the fitting parameters for $M_{\rm max}$ in Eq.~(1); $A, B, C$ are the fitting parameters for $R_{\rm eq, max}$ in Eq.~(2); $a, q, k$ are the fitting parameters for $I_{\rm max}$ in Eq.~(3). Taken from Ref.~\cite{ang16prd} and see there for details.}
\vspace*{-12pt}
\begin{center}
\def\temptablewidth{0.96\textwidth}
{\rule{\temptablewidth}{0.5pt}}
\begin{tabular*}{\temptablewidth}{@{\extracolsep{\fill}}cc|cccc|cc|ccc|ccc}
   \hline
   & & $P_{\rm K}$ & $I_{\rm K, max}$ & $M_{\rm TOV}$ & $~R_{\rm eq}~$  & $\alpha$ & $\beta$  & $A$ &  $B$ & $C$  & $a$ & $q$ & $k$ \\
   & EoS & (ms) & $(10^{45} {\rm g~cm}^2)$ & $(M_{\odot})$ & (km) & $(P^{-\beta})$  && $(P^{-B})$ && (km)&& (ms) & $(P^{-1})$ \\
   \hline
   & BCPM  & 0.5584 & 2.857 & 1.98 &  9.941  & 0.03859 & -2.651   & 0.7172 & -2.674 & 9.910 & 0.4509 & 0.3877 & 7.334  \\
NS & BSk20 & 0.5391 & 3.503 & 2.17 & 10.17 & 0.03587 & -2.675   & 0.6347 & -2.638 & 10.18  & 0.4714 & 0.4062 & 6.929  \\
   & BSk21 & 0.6021 & 4.368 & 2.28 & 11.08 & 0.04868 & -2.746   & 0.9429 & -2.696 & 11.03 & 0.4838 & 0.3500 & 7.085 \\
   & Shen & 0.7143 & 4.675 & 2.18 & 12.40 & 0.07657 & -2.738  & 1.393 & -3.431 & 12.47 & 0.4102 & 0.5725 & 8.644 \\
   \hline
   & CIDDM & 0.8326 & 8.645 & 2.09 & 12.43 & 0.16146 & -4.932   & 2.583 & -5.223 & 12.75 & 0.4433 & 0.8079 & 80.76  \\
QS & CDDM1 & 0.9960 & 11.67 & 2.21 & 13.99 & 0.39154 & -4.999   & 7.920 & -5.322 & 14.32 & 0.4253 & 0.9608 & 57.94  \\
   & CDDM2 & 1.1249 & 16.34 & 2.45 & 15.76 & 0.74477 & -5.175   & 17.27 & -5.479 & 16.13 & 0.4205 & 1.087 & 55.14 \\
      \hline
\end{tabular*}
      {\rule{\temptablewidth}{0.5pt}}
\end{center}
\end{table*}

While a previous analysis shows a qualitative consistency between this suggestion and the $Swift$ SGRB data, the distribution of observed break time $t_{\rm b}$ is much narrower than the distribution of the collapse time of supramassive NSs for the several NS EoSs investigated. We here include developed QS EoSs (labelled as CIDDM, CDDM1, CDDM2), as well as recently-constructed ¡°unified¡± NS EoSs (BCPM, BSk20, BSk21, Shen). The corresponding star properties are collected in the first four rows of Table~2. Rotating configurations ($M_{\rm max}, R_{\rm eq}, I_{\rm max}$) from the $rns$ code can be fitted well as a function of the spin period ($P$) (in millisecond) as follows:
\begin{eqnarray}
&& \frac{M_{\rm max}}{M_{\odot}} = \frac{M_{\rm TOV}}{M_{\odot}} \left[ 1 + \alpha~\left(\frac{P}{\rm ms}\right)^{\beta}\right]; \\
&& \frac{R_{\rm eq, max}}{\rm km} = C + A~\left(\frac{P}{\rm ms}\right)^B; \\
&& \frac{I_{\rm max}}{10^{45}{\rm g~cm^2}} = \frac{M_{\rm max}}{M_{\odot}} ~ \left(\frac{R_{\rm eq}}{\rm km}\right)^2 \frac{a}{1 + e^{-k~\left(\frac{P}{\rm ms} - q\right)}},
\end{eqnarray}
where the parameters ($\alpha, \beta, A, B, C, a, q, k$) are collected in the last eight columns of Table 2.

\begin{table*}
\tabcolsep 1pt
\caption{Simulated parameter ranges for supramassive NS/QS properties from the {\em Swift} internal plateaus sample with EoS models (BSk20, BSk21, Shen, CIDDM, CDDM1, CDDM2). Here $\epsilon$, $P_i$, $B_p$, $\eta$ are the ellipticity, the initial spin period, the surface dipole magnetic field, and the radiation efficiency, respectively. $P_{\rm best} (t_b)$ in the last column is the best $P$ value only for the $t_b$ distribution. Taken from Ref.~\cite{ang16prd} and see there for details.}
\vspace*{-12pt}
\begin{center}
\def\temptablewidth{0.98\textwidth}
{\rule{\temptablewidth}{0.5pt}}
\begin{tabular*}{\temptablewidth}{@{\extracolsep{\fill}}c|c|c|c|c|c}
\hline
EoS & $\epsilon$ & $P_i~(\rm ms)$ &$ B_p~(G)$ & $\eta$ & $P_{\rm best}~(t_{b})$\\
\hline
BSk20 & $0.002$ &   $0.70-0.75$   &  $N(\mu_{\rm Bp}=10^{14.8-15.4}, \sigma_{\rm Bp}\leq0.2)$   & $0.5-1$    & 0.20\\
BSk21 & $0.002$ &   $0.60-0.80$   &  $N(\mu_{\rm Bp}=10^{14.7-15.1}, \sigma_{\rm Bp}\leq0.2)$   & $0.7-1$    & 0.29 \\
Shen & $0.002-0.003$ &   $0.70-0.90$   &  $N(\mu_{\rm Bp}=10^{14.6-15.0}, \sigma_{\rm Bp}\leq0.2)$   & $0.5-1$    & 0.41 \\
\hline
CIDDM & $0.001$ &   $0.95-1.05$   &  $N(\mu_{\rm Bp}=10^{14.8-15.4}, \sigma_{\rm Bp}\leq0.2)$   & $0.5-1$    & 0.44\\
CDDM1 & $0.002-0.003$ &   $1.00-1.40$   &  $N(\mu_{\rm Bp}=10^{14.7-15.1}, \sigma_{\rm Bp}\leq0.3)$   & $0.5-1$    & 0.65 \\
CDDM2 & $0.004-0.007$ &   $1.10-1.70$   &  $N(\mu_{\rm Bp}=10^{14.8-15.3}, \sigma_{\rm Bp}\leq0.4)$   & $0.5-1$    & 0.84\\
\hline
\end{tabular*}
      {\rule{\temptablewidth}{0.5pt}}
\end{center}
\end{table*}

Before comparing our results with detailed SGRB observations, it is necessary to first check if the chosen NS/QS EoSs could reproduce the current fraction of supramassive stars, $\sim 22\%$, based on the mass distribution observation of Galactic NS-NS binary systems.   The details are in Ref.~\cite{ang16prd} and the results are shown in Fig.~5. One can see that all except the BCPM NS EoS can reproduce the $22\%$ fraction constraint (with slightly different required $P_i$). In the following we omit the BCPM EoS. The allowed static gravitational maximum mass $M_{\rm TOV}$ is then in the range of $2.17M_{\odot} - 2.45M_{\odot}$.

\begin{figure}[h]
\includegraphics[width=20pc]{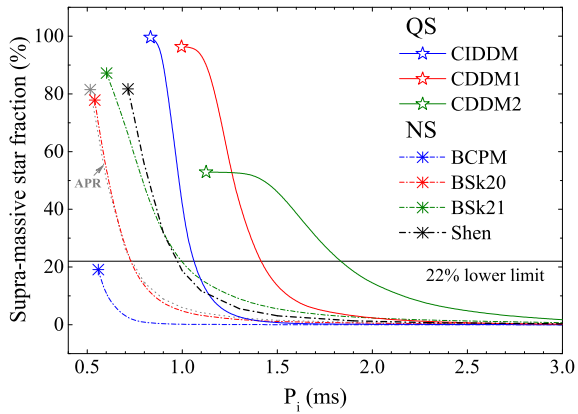}\hspace{2pc}%
\begin{minipage}[b]{14pc}
\caption{Theoretical estimations of the supramassive star fraction based on four cases of unified NS EoSs and three cases of QS EoSs, as compared with the observed $22\%$ constrain. Previous calculations using the APR NS EoS model~\cite{apr} are also shown for comparison. Taken from Ref.~\cite{ang16prd} and see there for details.}
\end{minipage}
\end{figure}

We finally apply our previous Monte Carlo simulations to the new EoSs for both NSs and QSs, trying to simultaneously reproduce all three distributions ($t_b$, $L_b$, $E_{\rm total}$) available from the observational data and revealing the post-merger supramassive stars' physics. The results are shown in Table 3 and Fig.~6.

In Table 3, we list the constrained ranges for the NSs'(QSs') parameters: an ellipticity $\epsilon$ as low as $0.002$ ($0.001$), an initial spin period $P_i$ commonly close to the Keplerian limit $P_{\rm K}$, a surface dipole magnetic field of $B_p \sim 10^{15}$ G [satisfying normal distributions $N(\mu_{\rm Bp}, \sigma_{\rm Bp})$], and an efficiency of $\eta = 0.5 - 1$ related to the conversion of the dipole spin-down luminosity to the observed X-ray luminosity. In the last column we show $P_{\rm best} (t_b)$, the best values only for the $t_b$ distribution. It is clear that the KS test for the $t_b$ distribution is indeed improved. In particular, as one can see from Fig.~6, the $t_b$ distributions in the QS scenarios are more concentrated, which provide a better agreement with the observed ones. Some quantitative differences would exist for other QS model EoSs, but same conclusions should hold because of the similar QS rotational properties obtained (more pronounced increases of $M_{\rm max}$ with frequency than those with the NS ones: $\sim40\%$ vs. $18-20\%$ in the present calculation). We therefore argue that a supramassive QS is favored than a supramassive NS to serve as the central engine of SGRBs with internal plateaus. NS-NS mergers are a plausible location for quark deconfinement and the formation of QSs.

\begin{figure}[h]
\includegraphics[width=20pc]{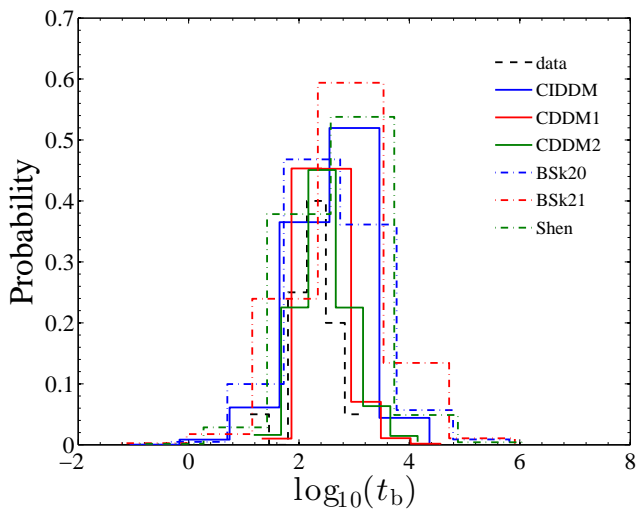}\hspace{2pc}%
\begin{minipage}[b]{14pc}\caption{Simulated collapse time distributions with three unified NS EoSs and three QS EoSs, as compared with the observed one (dashed curve). Taken from Ref.~\cite{ang16prd} and see there for details.}
\end{minipage}
\end{figure}

\section{Summary and future perspectives}

In the present proceeding we show our three recent calculations on rotating compact stars, and confront them with available/upcoming accurate astrophysical observations, including pulsar glitch, MoI, and SQRB.

In terms of future methods, useful constrains could be possible by combining gravitational wave and electromagnetic observations from both coalescing NS-NS binaries and isolated NSs. It would be interesting and meaningful to study whether NS EoS could be determined from gravitational wave emission from elliptically deformed pulsars and NS-NS binaries.

\section*{Acknowledgments}
The work was supported by the National Natural Science Foundation of China (Nos 11405223, U1431107).

\end{document}